\documentclass[aps,pra,showpacs,twocolumn]{revtex4-1}

\usepackage{amsmath,amssymb,stmaryrd} 
\usepackage{natbib}

\usepackage[draft,marginclue,margin,author=]{fixme}
\fxsetup{theme=color,marginface=\small}

\usepackage{ifpdf} 
\ifpdf 
  \usepackage[pdftex,breaklinks=true]{hyperref}
  \usepackage{graphicx} 
\else  
  \usepackage[hypertex,breaklinks=true]{hyperref}
  \usepackage{graphicx} 
\fi

\newcommand{\Tr}{\mathop{\mathrm{Tr}}\nolimits}
 \renewcommand{\vec}[1]{\boldsymbol{#1}}
 
\renewcommand{\Im}{\mathop{\mathrm{Im}}\nolimits} 

\begin{document}

\title{Exact path-integral evaluation of locally interacting systems: The subtlety of operator ordering}

\author{Nobuhiko Taniguchi}
\email{taniguchi.n.gf@u.tsukuba.ac.jp}
\affiliation{Physics Division, Faculty of Pure and Applied Sciences,
  University of Tsukuba, Tennodai Tsukuba 305-8571, Japan}

\date{\today}

\begin{abstract}
  We discuss how one calculates the coherent path integrals for
  locally interacting systems, where some inconsistencies with exact
  results have been reported previously. It is shown that the operator
  ordering subtlety that is hidden in the local interaction term
  modifies the Hubbard-Stratonovich transformation in the continuous
  time formulation, and it helps reproduce known results by the
  operator method.  We also demonstrate that many-body effects in the
  strong interaction limit can be well characterized by the
  free-particle theory that is subject to annealed random potentials
  and dynamical gauge (or phase) fields.  The present treatment
  expands the conventional paradigm of the one-particle description,
  and it provides a simple, viable picture for strongly correlated
  materials of either bosonic or fermionic systems.
\end{abstract}




\maketitle%


\section{Introduction} 

The path integral formulation~\cite{FeynmanBook65,KleinertBook09} has
been widely used in many areas of physics and has now become an
indispensable tool in formulating, investigating and understanding
quantum physics.  Its variant, the coherent-state path
integral~\cite{KlauderBook85,Zhang90}, is particularly useful and
versatile for analyzing quantum many-body theories where the
Hamiltonian is expressed in normal-ordered products of creation
and annihilation operators~\cite{AltlandBook10}. It helps us 
handle either bosons or fermions, perform a perturbational expansion,
treat nonperturbative contributions like topological effects,
and grasp relevant physics intuitively.

The downside of the path integral approach is that its direct evaluation
tends to demand more effort than that of the operator method.  
Even for noninteracting quadratic Hamiltonians, great care is needed
to tackle the operator ordering subtlety or a seemingly divergent
determinant.  
The situation gets exacerbated for interacting systems, even for the
simplest possible interacting system, namely the one-site Bose-Hubbard
model.  When one uses the time-continuous coherent-state path integral
to evaluate, say, $\Tr [e^{-\beta U c^{\dagger} c^{\dagger} c c/2}]$
with a single bosonic field, one may well be deceived into reaching
the wrong answer $\sum_{n=0}^{\infty} e^{-\beta U n^{2}/2}$, instead
of the correct one
$\sum_{n=0}^{\infty} e^{-\beta U n(n-1)/2}$~\cite{Wilson11,Kordas14}.
The form of discrepancy strongly suggests that the approach may be plagued by
some operator ordering subtlety that the quartic term may have.  The
same problem prevails in many-particle systems with local interaction,
either of bosons or of fermions.

In order to remedy this embarrassing situation, \citet{Wilson11}
surmised that an additional correction is present in the
representation of the normal-ordered interaction in a way to reproduce
the exact result.  \citet{Kordas14} subsequently proposed a possible,
consistent redefinition of the coherent-state path integral
formulation that successfully reproduces the correct results of the
one-site Bose-Hubbard model.
The scheme is non-standard, though. Starting with a normal-ordered
Hamiltonian, they expressed normal-ordered operators in terms of the
coordinate-momentum representation (or the Weyl symbol).  The
procedure is inconvenient and nontrivial when one tries to treat 
many-particle bosonic systems whose degree of freedom is large,
not to mention systems that involve many fermions.
Because the many-body Hamiltonian is normal-ordered, complying with
the standard coherent-state formulation has a clear benefit.  It is
worth understanding what goes wrong in its conventional treatment and
finding a simple, reliable way of reaching correct results.

\paragraph*{Purpose}

In this paper, we reexamine the coherent-state path integral for
locally interacting many-body systems where constituent particles are
either bosons or fermions.  Like the one-site Bose-Hubbard model, the
coherent-path integral seemingly fails to reproduce the exact results
of the partition function and Green function, if one uses the
conventional definition.  We scrutinize the evaluation process and
identify the cause in the operator ordering subtlety hidden in the
interaction term.  We find that circumventing that subtlety makes us
modify the Hubbard-Stratonovich (HS) transformation.  Accordingly, one
can readily reproduce the known exact results in the standard
definition of the coherent-state path integral.
Our discussion focuses on a simple type of local interaction defined
in Eq.~\eqref{eq:Hamiltonian}, but the same argument can straightforwardly
apply to a more general form of the interaction among mutually
commuting operators, while treating the interaction between mutually non-commuting operators is nontrivial (see Appendix~\ref{app:extension-of-HS}). 

Locally interacting models can be viewed as the strong interaction
limit of correlated materials where the interaction is much greater
than the band width so that each site is effectively isolated.
Charge-blocking, many-body Mott physics dominates, and the
one-particle picture gets inappropriate.  Propagating degrees of
freedom responsible for such dynamical gap is
elusive~\cite{Phillips10}.
In the process of evaluating the path integral, we will encounter a
certain free-particle theory that is subject to dynamical phase fields
and random potentials. This supplement to the free-particle theory is
of great interest because it tells how the free-particle theory can
accommodate non-perturbative many-body correlation.
One can describe strongly correlated materials by using emergent gauge
field~\cite{Lee06}.  Because of charge blocking, the phase degree
fluctuates dynamically far beyond Gaussian, and so does the gauge
field, which is the time derivative of the phase field.  In this
respect, one may regard the present calculation as a concrete example
of how to analyze dynamical fluctuations of the emergent gauge field
in the strong interaction limit.

\section{Locally interacting systems}

\subsection{Model}

We consider a multi-level (or multi-site) system of bosonic or
fermionic particles $(\psi_{\alpha},\psi_{\alpha}^{\dagger})$, which
interact locally.  The Hamiltonian is given by
\begin{align}
& \hat{H} = \sum_{\alpha} \epsilon_{\alpha} \hat{n}_{\alpha} 
+ \frac{U}{2} \hat{N}(\hat{N}-1),
\label{eq:Hamiltonian}
\end{align}
where the label $\alpha$ refer to levels and/or spins, and
$\hat{N} = \sum_{\alpha} \hat{n}_{\alpha} = \sum_{\alpha}
\psi_{\alpha}^{\dagger} \psi_{\alpha}$ is the total number operator.
In spite of the interaction being present, one can exactly calculate
thermodynamics and various Green functions by help of the operator
method (see Appendix~\ref{app:operator-method}).  Yet, with the
coherent-state path integral, one must be cautious to reach those
results.

\subsection{Subtlety disclosed}

We start by revealing a subtlety hidden in the standard
manipulation of the coherent-path integral. Taking the grand
partition function
$\Xi_{U}(\mu)=\Tr [ e^{-\beta (\hat{H}-\mu \hat{N})}]$,
we can establish the exact identity between
$\Xi_{U}(\mu)$ and the noninteracting counterpart $\Xi_{0}(\mu)$: 
\begin{align}
  & \Xi_{U}(\mu) = 
    \int^{\infty}_{-\infty}\!\! d[\tilde{\varphi}]\, e^{-
\beta \frac{\tilde{\varphi}^{2}}{2U}}\, 
    \Xi_{0}(\mu + \tfrac{U}{2} - i\tilde{\varphi}),
 \label{eq:Xi-identity}
\end{align}
which is derived in Eq.~\eqref{eq:XiU-Xi0}.  
Here $\tilde{\varphi}$ denotes a time-independent Gaussian variable
with variance $U/\beta$ and the measure
$d[\tilde{\varphi}]=\sqrt{\beta/2\pi U}d\tilde{\varphi}$ includes the
normalization. Relation \eqref{eq:Xi-identity} holds for either bosons
or fermions. 
One can see Eq.~\eqref{eq:Xi-identity} come from the operator
identity [see Eq.~\eqref{eq:op-identity-detail}],
\begin{align}
& e^{-\beta \frac{U}{2} \hat{N}^{2}}
= \int^{\infty}_{-\infty} d[\tilde{\varphi}]\,
  e^{-\beta \frac{\tilde{\varphi}^{2}}{2U} - i\beta \tilde{\varphi}
  \hat{N}}.  
\label{eq:op-identity}
\end{align}
The formula can be viewed as an operative version of the HS
transformation.  An important observation is that when we take the
coherent-path integral representation of Eq.~\eqref{eq:op-identity},
it contradicts the standard form of the HS transformation. 
Indeed, the decomposition concerning $e^{-\beta \hat{H}}$ becomes
(see Appendix~\ref{app:detailed-derivation} for the derivation)
\begin{subequations}
\label{eq:thermal-HS-decoupling}
\begin{align}
& \int \mathcal{D}[\psi,\bar{\psi}]\,
e^{- \mathcal{S}/\hbar}
= \int \mathcal{D}[\tilde{\phi}] \mathcal{D}[\psi,\bar{\psi}]\,
e^{- (\mathcal{S}_{e}+\mathcal{S}_{\phi})/\hbar},
\end{align}
where the Euclidean actions $\mathcal{S}$ and $\mathcal{S}_{e,\phi}$
are defined by
\begin{align}
  & \mathcal{S} = \sum_{\alpha,\beta}\int^{\beta\hbar}_{0}\!\! d\tau\,
\bar{\psi}_{\alpha} \left[ \left( \hbar  \partial_{\tau} 
    + \epsilon_{\alpha} \right) \delta_{\alpha \beta} +
    \frac{U}{2} \bar{\psi}_{\beta} \psi_{\beta} \right]
    \psi_{\alpha}, \\
& \mathcal{S}_{e} 
= \int^{\beta\hbar}_{0} d\tau \sum_{\alpha}
                \bar{\psi}_{\alpha} \left( \hbar \partial_{\tau} +
                \epsilon_{\alpha} - \frac{U}{2} + i\tilde{\phi} 
                \right) \psi_{\alpha}, \\
& \mathcal{S}_{\phi} =  \int^{\beta\hbar}_{0} d\tau\, 
\frac{\tilde{\phi}^{2}}{2U}.  
\end{align}
\end{subequations}
The above formula differs from the standard HS formula by the presence
of $-U/2$ in $\mathcal{S}_{e}$.  It is caused by circumventing the
operator ordering subtlety hidden in the standard derivation of the HS
transformation (see Appendix~\ref{app:subtlety-of-HS}), and tells us
to modify the HS transformation, when we comply with the standard
definition of the coherent state path integral.
With the modified representation of the interaction, we can readily
evaluate the path integral expression of $\Xi_{U}(\mu)$ by following
each step of Appendix~\ref{app:detailed-derivation} reversely.

In addition to the operator ordering subtlety, the HS transformation
has been known to be plagued by the ambiguity in selecting relevant
channels~\cite{[{See }][{}]Kleinert11,*[{}][{ \S
    14.15.}]KleinertBook16}.  When truncating relevant fluctuations,
it causes a serious problem that might give a different physical
result.  In the present treatment, however, we don't have such a
problem, because we carry out the complete integration of the
auxiliary fields without any approximation, thanks to the gauge
transformation.
Moreover, perturbative treatment often brings divergent contributions
due to interaction, and it therefore needs an additional renormalization
procedure.  This is not the case here, because the knowledge of
$\Xi_{0}(\mu)$ is sufficient to calculate $\Xi_{U}(\mu)$ exactly.

\subsection{Green functions}

We now turn our attention to evaluating various one-particle Green
functions of locally interacting systems.  Below, we use the
closed-time path integral
formalism~\cite{AltlandBook10,KadanoffBook62,Keldysh65,RammerBook07,KamenevBook11}
to formulate real-time correlation functions.  We show how we can
exactly evaluate those path integral representations by using a gauge
transformation
technique~\cite{Kamenev96,Kleinert97,Florens02,Florens03,Sedlmayr06}.
Such approach was undertaken in~\cite{Sedlmayr06} to investigate the
tunneling density of states at Coulomb-blockade peaks of fermionic
locally interacting systems, but its exposition is too succinct to
clarify the subtlety of the coherent-state path integrals. We
demonstrate how the modified HS transformation
[Eq.~\eqref{eq:HS-decoupling} below], which extends
Eqs.~\eqref{eq:thermal-HS-decoupling} to include real-time paths,
enables us to evaluate them.  Later in
Sec. III, we will show that they are identical to what are calculated
by the operator method.
Moreover, we find that Green functions for a locally interacting
system can be connected and determined by the knowledge of
noninteracting systems, like the grand partition function [see
Eq.~\eqref{eq:G-U} or \eqref{eq:Glesser-U} below].

\begin{figure}
  \centering
  \includegraphics[width=0.85\linewidth]{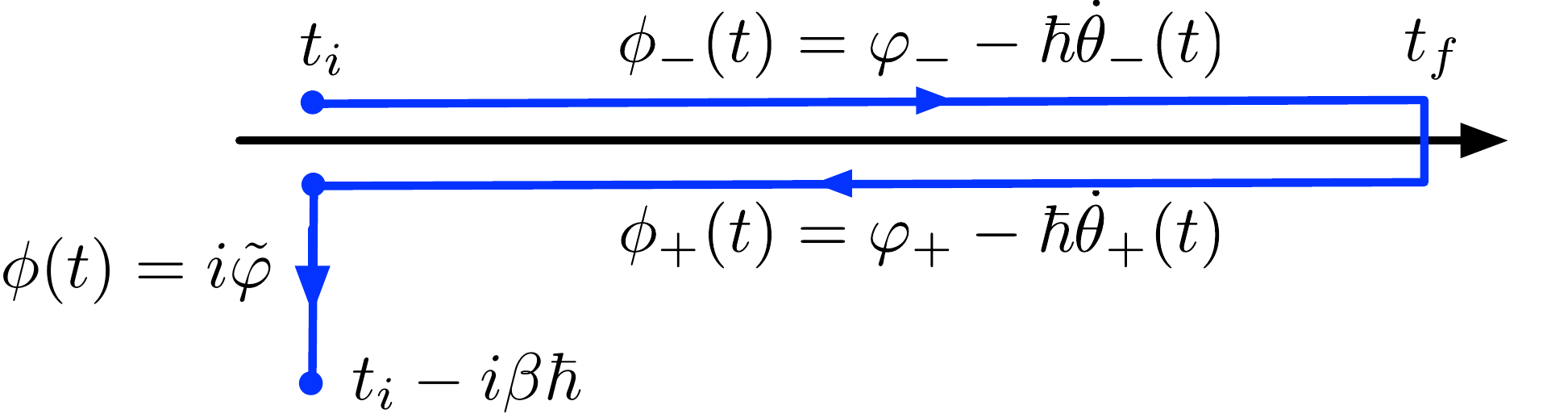}
  \caption{The Keldysh contour is composed of three segments: the
    forward-going $t_{i} \to t_{f}$, the backward-going
    $t_{f}\to t_{i}$, and the thermal one
    $t_{i} \to t_{i}-i\beta\hbar$. The arrows specify the contour
    ordering. Time arguments of Green functions reside between $t_{i}$
    and $t_{f}$, and the infinite limit of the time span
    $\Delta t = t_{f}-t_{i} \to \infty$ is taken.  On each segment,
    one decompose the auxiliary field $\phi$ into the static zero-mode
    $\varphi$ and the dynamical non-zero mode $\theta$. And one can
    safely gauge away the non-zero mode on the thermal segment.  }
  \label{fig:Keldysh-path}
\end{figure}

We define four types of real-time Green functions, 
\begin{align}
&\begin{pmatrix}
 G_{\alpha} (t,0)& G^{<}_{\alpha}(t,0) \\ 
  G^{>}_{\alpha}(t,0) & \tilde{G}_{\alpha}(t,0)
\end{pmatrix}
= \frac{1}{i\hbar} 
\begin{pmatrix}
 \langle T \psi_{\alpha}(t) \psi_{\alpha}^{\dagger} \rangle 
  & \pm \langle\psi_{\alpha}^{\dagger} \psi_{\alpha} (t) \rangle \\ 
  \langle \psi_{\alpha} (t) \psi_{\alpha}^{\dagger}\rangle 
  & \langle \tilde{T} \psi_{\alpha} (t) \psi_{\alpha}^{\dagger}\rangle
\end{pmatrix},
\end{align}
where $\pm$ refers to bosonic or fermionic systems, and
$\langle \cdots \rangle$ is the thermal average specified by chemical
potential $\mu$ and the inverse temperature $\beta$.  The operator $T$
is the time-ordering operator, while $\tilde{T}$ is the
anti-time-ordering one.  We can compactly write them by the contouring-ordering operator $T_{c}$ along the Keldysh
path $\int_{K}$ as
\begin{align}
& G_{\alpha} (t_{1},t_{2}) = \frac{1}{i\hbar} \left\langle T_{c} \psi_{\alpha}
  (t_{1}) \, \psi_{\alpha}^{\dagger} (t_{2}) \right\rangle,
\\ & \quad
= \frac{1}{i\hbar \Xi_{U}} \int\!\!
  \mathcal{D}[\psi,\bar{\psi}]\, 
\psi_{\alpha}(1) \bar{\psi}_{\alpha}(2) \, e^{\frac{i}{\hbar} S},
\end{align}
where the path is composed of three segments (see
Fig.~\ref{fig:Keldysh-path}): the forward-going (denoting $-$) from
the initial time $t_{i}$ to the final time $t_{f}$, the backward-going
(denoting $+$) from $t_{f}$ to $t_{i}$, and the thermal one from
$t_{i}$ to $t_{i}- i\beta \hbar$.
Since the interaction $U\hat{N}(\hat{N}-1)/2$ is normal-ordered, the action
$S$ that appears in the coherent-state path integral becomes
\begin{align}
& S = \int_{K} \sum_{\alpha,\beta} \bar{\psi}_{\alpha} \Big[ 
( i\hbar \partial_{t} - \epsilon_{\alpha}) \delta_{\alpha\beta}
- \frac{U}{2} \bar{\psi}_{\beta} \psi_{\beta} \Big]
  \psi_{\alpha}.
\end{align}
The next step is crucial: we decompose the interaction term via the
modified HS transformation along the Keldysh path.  The transformation is
\begin{align}
& e^{-\frac{i}{\hbar} \int_{K} \frac{U}{2} N^{2}(t)}
= \int \mathcal{D}[\phi]\, e^{\frac{i}{\hbar} (S_{\phi} + S_{e})}. 
\label{eq:HS-decoupling}
\end{align}
where
\begin{align}
& S_{\phi} = \int_{K} \frac{\phi^{2}(t)}{2U}, \\
& S_{e} = \int_{K}  \sum_{\alpha}
  \bar{\psi}_{\alpha} \Big[ 
  i\hbar \partial_{t} - \epsilon_{\alpha} - \phi(t)+ U/2
  \Big] \psi_{\alpha}.
\end{align}
The term $U/2$ is mandatory in $S_{e}$, as in Eq.~\eqref{eq:thermal-HS-decoupling}.  

After we have managed the operator ordering subtlety, we may follow
the observation
in~\cite{Kamenev96,Kleinert97,Florens02,Florens03,Sedlmayr06} to
employ the local gauge transformation to absorb most of the effect of
$\phi(t)$. To make this work, however, we have to carefully specify
the boundary condition: the periodicity of $\phi(t)$
must be imposed on \emph{each} of the three segments of the Keldysh
path, to ensure new field operators ($\Psi_{\alpha}$ below) to remain
canonical.
We then decompose $\phi$-fields on each segment into the
static zero-modes $\varphi=(\varphi_{\mp},i\tilde{\varphi})$, and the
dynamical phase fields $\theta(t) = \theta_{\mp}(t)$ satisfying the periodic boundary condition: 
\begin{align}
  & \psi_{\alpha}(t) = e^{i\theta (t)} \Psi_{\alpha}(t); \quad
 \phi(t) = \varphi - \hbar \dot{\theta}(t).
  \label{eq:gauge-transformation}
\end{align}
One can safely gauge away the dynamical field on the thermal segment. 
Now the action becomes 
\begin{align}
& S_{e} = \int_{K} \sum_{\alpha} 
  \bar{\Psi}_{\alpha} \left[ 
  i\hbar \partial_{t} - \epsilon_{\alpha} - \varphi +U/2
  \right] \Psi_{\alpha}, \\
& S_{\phi} = \frac{i\beta \hbar}{2U}\tilde{\varphi}^{2}
+ \frac{\Delta t}{2U} \left( \varphi_{-}^{2} - \varphi_{+}^{2}
  \right) +  \int_{K} \frac{\hbar^{2} \dot{\theta}^{2}}{2U},
\end{align}
with $t_{f}-t_{i} = \Delta t$. 

The dynamical phase fields may be regarded as compact $U(1)$ gauge
fields that commonly emerge in strongly correlated
matter~\cite{Lee06}.  One may examine nonperturbative correlation
effect by studying nontrivial field configurations that carry finite
winding numbers.
In the present approach, the dynamical fields $\theta_{\mp}$ describe
the fluctuating part on top of nontrivial field configurations,
while $\varphi = (\varphi_{\pm}, i\tilde{\varphi}$ affects
the thermodynamics and its dynamics nonperturbatively.

The result of the $\Psi$-integral can be written by the Green
functions multiplied by the grand partition function of the
noninteracting particles.  We still need to complete the $\varphi$-
and $\theta$-integrals, but in isolated systems here, those integrals
are found to be decoupled~\cite{Kamenev96,Kleinert97,Sedlmayr06}.
Symbolically, one can write the result as
\begin{align}
  & G_{\alpha} (1,2) = \frac{1}{\Xi_{U}}\Big\langle \Xi^{\varphi}
    G_{\alpha}^{\varphi} (1,2) \Big\rangle_{\varphi} 
\Big\langle
    e^{i\theta(1)} e^{-i\theta(2)} \Big\rangle_{\theta},
\label{eq:G-U}
\end{align}
where $\langle \cdots \rangle_{\varphi}$ refers to the Gaussian
average over the three static Gaussian variables
$(\varphi_{\mp}, \tilde{\varphi})$, while
$\langle \cdots \rangle_{\theta}$, to the path integration over
dynamical $\theta$. 
The explicit forms of $\Xi^{\varphi}$ and $G_{\alpha}^{\varphi}$ are nothing but the noninteracting ones, $\Xi_{0}$ and $G_{0,\alpha}$, where
\begin{align}
& \Xi_{0}^{\varphi} = \Xi_{0} (\{\epsilon_{\alpha}^{\varphi},
  \mu^{\varphi}\}) = \prod_{\alpha} \left[ 1 \mp e^{-\beta
  (\epsilon_{\alpha}^{\varphi} - \mu^{\varphi})} \right]^{\mp 1}, \\ 
& G_{\alpha}^{\varphi}(t,0) = G_{0,\alpha}(t; \{
  \epsilon_{\alpha}^{\varphi}, 
  \mu^{\varphi} \}),
\end{align}
 with incorporating $\varphi$-dependence by shifting 
 $\epsilon_{\alpha}$ and $\mu$ by
\begin{align}
& \epsilon_{\alpha}^{\varphi} = \epsilon_{\alpha} - \frac{U}{2} +
  \varphi_{c}; \quad
 \mu^{\varphi} = \mu - i\tilde{\varphi} + \varphi_{c} - \frac{i
  \Delta t}{\beta \hbar} \varphi_{q}.
\label{eq:replacing-rules}
\end{align}
Here the convention $\varphi_{c} = (\varphi_{-}+\varphi_{+})/2$ and
$\varphi_{q} = \varphi_{-} - \varphi_{+}$ is used. 
When we recover the Keldysh structure, the part
$\langle e^{i\theta(1)} e^{-i\theta(2)} \rangle_{\theta}$ in
Eq.~\eqref{eq:G-U} means the contour-ordered vertex correlator.  We
can calculate it as the action regarding $\theta$ is free-particle 
with mass $U/\hbar^{2}$. 
Though local fluctuation $\big\langle \theta^{2} (t) \big\rangle$
diverges, it is finite and equal to 
\begin{align}
& \left\langle T_{c}\, e^{-i\theta(t)} e^{i\theta(0)} \right\rangle_{\theta}
= \begin{pmatrix} 
  e^{-\frac{iU}{2\hbar}|t|} & e^{\frac{iU}{2\hbar}t} \\ 
  e^{-\frac{iU}{2\hbar}t} & e^{\frac{iU}{2\hbar}|t|}
 \end{pmatrix}.
\end{align}
Combining all the above, we can evaluate exactly all one-particle
Green functions for locally interacting systems.

Let us briefly illustrate how it operates in practice.  The lesser
component of Eq.~\eqref{eq:G-U} gives
\begin{align}
& G_{\alpha}^{<} (t,0) = \frac{1}{\Xi_{U}}
\left\langle \Xi_{0}^{\varphi} G^{\varphi,<}_{\alpha}(t,0)
  \right\rangle_{\varphi} \;  e^{\frac{iU}{2\hbar} t},
\label{eq:Glesser-U}
\end{align}
and the noninteracting lesser Green function is
\begin{align}
& G^{\varphi,<}_{\alpha} (t,0) 
= \pm \frac{e^{-\frac{i}{\hbar} \epsilon_{\alpha}^{\varphi}
  t}}{i\hbar} n_{\alpha}^{\varphi}.
\end{align}
The occupation $n_{\alpha}^{\varphi}=\langle \hat{n}_{\alpha}\rangle$
has to be determined by the partition function $\Xi^{\varphi}$ via the
standard relation, 
\begin{align}
& n_{\alpha}^{\varphi} = - \frac{1}{\beta} \frac{\partial}{\partial
  \epsilon_{\alpha}} \ln \Xi_{0}^{\varphi}.  
\end{align}
It means that $G_{\alpha}^{<}(t,0)$ of locally interacting systems is
expressed in a form of the \emph{annealed average} over three random (static)
Gaussian variables $\varphi=(\varphi_{\mp}, \tilde{\varphi})$:
\begin{align}
& G_{\alpha}^{<} (t,0) 
= \mp \frac{1}{i\hbar \Xi_{U}}\,
  \left\langle e^{-\frac{i}{\hbar}
  (\epsilon_{\alpha}^{\varphi}- \frac{U}{2})t}\, \frac{\partial 
  \Xi_{0}^{\varphi}}{\beta\partial \epsilon_{\alpha}}
  \right\rangle_{\varphi}. 
\label{eq:G-lesser-path-int}
\end{align}
We can likewise find the greater Green
function, 
\begin{align}
& G_{\alpha}^{>}(t,0) 
= \frac{1}{i\hbar \Xi_{U}} 
  \left\langle e^{-\frac{i}{\hbar}
  (\epsilon_{\alpha}^{\varphi}+ \frac{U}{2})t}\, 
\left[ \Xi_{0}^{\varphi} \mp \frac{\partial 
  \Xi_{0}^{\varphi}}{\beta\partial \epsilon_{\alpha}} \right]
  \right\rangle_{\varphi}. 
\label{eq:G-greater-path-int}
\end{align}
From these results of $G_{\alpha}^{<}$ and $G_{\alpha}^{>}$, we can construct all the other one-particle Green functions. 

\section{Equivalence to the operator method}

We now check that the results Eqs.~\eqref{eq:G-lesser-path-int} and
\eqref{eq:G-greater-path-int} actually reproduce the Green functions
evaluated by the operator method in
Appendix~\ref{app:operator-method}.
To see it, we expand $\Xi_{0}^{\varphi}$ in terms of the canonical
partition function $Z_{N}$ of non-shifting levels $\epsilon_{\alpha}$,
\begin{align}
& \Xi_{0}^{\varphi} = \sum_{N=0}^{\infty} Z_{N} \, e^{N \beta
  (\mu + \frac{U}{2}- i\tilde{\varphi} -
  \frac{i\Delta t}{\beta \hbar} \varphi_{q})}.
\end{align}
We find that the integration over $\varphi_{q}$ simply enforces
$\varphi_{c}/U$ to non-negative integers $N$ in the limit of
$\Delta t \to\infty$.
\begin{align}
& \int d[\varphi_{q}] \, e^{\frac{i \Delta t}{\hbar U} \varphi_{c}
  \varphi_{q}}   e^{-\frac{i\Delta t}{\hbar} \varphi_{q} N} 
= \delta(\varphi_{c} -U N).
\label{eq:phi-c-locking}
\end{align}
Accordingly, we may say that $\varphi_{c}/U$ plays a role of winding
numbers of the emergent compact gauge field configuration; a naive
saddle-point (or Hartree-Fock) approximation regarding $\varphi$
misses such nonperturbative contribution. We need to take account of
all the contribution of $N$ on principle (see Ref.~\cite{Sedlmayr06}
for its implication on the tunneling density of states). 
By completing the remaining Gaussian average over $\tilde{\varphi}$, we
organize the result as
\begin{subequations}
\begin{align}
& G_{\alpha}^{<} (t,0) 
= \frac{\pm 1}{i\hbar}\sum_{N=0}^{\infty} e^{-\frac{i}{\hbar}
     [\epsilon_{\alpha}+U(N-1)]t}\,  n_{\alpha|N}, \\
& G_{\alpha}^{<} (\varepsilon) 
= \mp 2i\pi \sum_{N=0}^{\infty} n_{\alpha|N} \, \delta \big( \varepsilon -
  \epsilon_{\alpha} -  U(N-1) \big).
\end{align}
\label{eq:G-lesser-by-n}
\end{subequations}
Here we have introduced the quantity $n_{\alpha|N}$,  the ``fractional 
parentage'' of the occupation number onto the fixed $N$.  It is defined by 
\begin{align}
& n_{\alpha|N}
= -\frac{1}{\beta\Xi_{U}} \frac{\partial Z_{N}}{\partial
  \epsilon_{\alpha}} e^{\beta N\mu - \beta\frac{U}{2}N(N-1)},
\end{align}
and satisfies
$\langle \hat{n}_{\alpha} \rangle = \sum_{N=0}^{\infty} n_{\alpha|N}$. 
Similarly, we find the greater Green function to be
\begin{subequations}
\label{eq:G-greater-by-p}
\begin{align}
& G_{\alpha}^{>}(t,0) = \frac{1}{i\hbar} \sum_{N=0}^{\infty} e^{-\frac{i}{\hbar}
  (\epsilon_{\alpha} + UN)t} p_{\alpha|N}, \\
& G_{\alpha}^{>}(\varepsilon) = -2i\pi \sum_{N=0}^{\infty} p_{\alpha|N}\;
  \delta(\varepsilon - \epsilon_{\alpha} - UN),
\end{align}
\end{subequations}
by using $p_{\alpha|N}$, the fractional parentage of the hole
occupation onto a fixed $N$, defined by 
\begin{align}
& p_{\alpha|N} = \frac{1}{\Xi_{U}}\left[ Z_{N} \mp \frac{1}{\beta}
  \frac{\partial Z_{N}}{\partial \epsilon_{\alpha}} \right] e^{\beta
  N\mu - \beta \frac{U}{2}N(N-1)}.
\end{align}
The spectral function $\rho_{\alpha}(\varepsilon)$ is
straightforwardly calculated as
\begin{align}
& \rho_{\alpha} (\varepsilon) = \sum_{N=0}^{\infty} \Big[ p_{\alpha|N}\,
  \delta(\varepsilon - \epsilon_{\alpha} - U N) 
\notag \\ & \qquad \qquad
\mp n_{\alpha|N}\,
  \delta(\varepsilon - \epsilon_{\alpha} - U(N-1))  \Big]. 
\label{eq:spectral-function}
\end{align}
In these forms, one can confirm the equivalence with the ones by the operator method in Appendix~\ref{app:operator-method}.

\section{Discussion} 

We have shown that we can treat a locally interacting system correctly
using the standard definition of the coherent-path integral. The
results are connected with their noninteracting counterpart. [See
Eqs.~\eqref{eq:Xi-identity} for the partition function, and
\eqref{eq:G-lesser-path-int}-\eqref{eq:G-greater-path-int} for Green
functions.]
The relation~\eqref{eq:Xi-identity} shows that the thermodynamics of a
locally interacting system is exactly equivalent to the annealed
average of the noninteracting Hamiltonian with random imaginary
potential $\tilde{\varphi}$.  Such simple correspondence, however,
cannot be held for Green functions
\eqref{eq:G-lesser-path-int}-\eqref{eq:G-greater-path-int} --- they
are still written by a free-particle model under the influence of
static random fields, as is seen in Eq.~\eqref{eq:G-U}, but we can
assign no single random Hamiltonian for its dynamics, because three
independent random variables are needed: $\varphi_{\mp}$ along the two
real-time paths and $\tilde{\varphi}$ on the thermal path.
We stress that this supplement to the free-particle theory can fully
capture various many-body characteristics like atomic correlations,
non-rigid bands, asymmetry of particle and hole excitations.  While a
spectral function in the conventional one-particle/quasiparticle
picture has only a single peak, the function
$\rho_{\alpha}(\varepsilon)$ of Eq.~\eqref{eq:spectral-function} has
multiple peaks with different weights at
$\varepsilon = \epsilon_{\alpha} + UN$. At those energies, the
retarded self-energy diverges and the retarded Green function vanishes, which
signals the demise of the quasiparticle picture~\cite{Phillips10}.

To treat non-perturbative many-body effect, it is important to take
account of two aspects: discreteness of the particle number and large
phase fluctuations beyond quadratic order. They are closely
related. We can implement discreteness of $N$ by compactifying the
conjugate phase $\Theta$ modulo $2\pi$ (satisfying
$[\hat{N}, \hat{\Theta}] = i$). Non-positive nature of $N$ makes
$\Theta$ non-Hermite~\cite{Lynch95}.
Since the phase $\Theta(t)$ couples linearly with $\dot{N}(t)$, we may
take the HS field $\phi(t)$ as $\phi(t) = \hbar
\dot{\Theta}(t)$. It means that we need to treat fluctuations of
$\phi(t)$ consistently by respecting such nontrivial nature of $\Theta$.
A common practice after introducing the HS field $\phi(t)$ is to
complete the quadratic integration over the field
$(\psi_{\alpha}, \bar{\psi}_{\alpha})$, and then to take the
saddle-point approximation regarding $\phi$.  Assuming a uniform
solution $\phi(t) = \varphi_{sp}$, one can determine the
self-consistent saddle-point solution $\varphi_{sp}$ by the average
number $\langle \hat{N} \rangle = \varphi_{sp}/U$ in that
approximation.  This contrasts with the exact locking of
$\varphi_{c}/U$ to non-negative integers in 
Eq.~\eqref{eq:phi-c-locking}.  A physical picture given by the
saddle-point approximation is fundamentally wrong, having no dynamical
gap generation and retaining the non-interacting Fermi-Dirac form of
the occupation $\langle n_{\alpha} \rangle$.
We find the gauge transformation technique is effective to
incorporating many-body effects.  Without any additional ansatz of the
slave-particle, one can describe many-body charge-blocking physics.

In hindsight, it is because the local occupation number is conserved
that one can solve locally interacting systems exactly.  When we
couple a locally interacting system linearly with external
environments (reservoirs), the local occupation is no longer
conserved, and the integrals over $\varphi$ and $\theta$ are coupled
unlike Eq.~\eqref{eq:G-U}. It seems unlikely that we can complete the
remaining path integrals exactly.  Nevertheless, the present analysis
of path integrals provides a useful and systematic means to describe
the local strong correlation that perturbation theory cannot treat.
In a quantum dot coupled to the leads, two types of strongly
correlated phenomena are known to emerge: the Coulomb blockade (or
charge-blocking due to correlation) and the Kondo
physics~\cite{Aleiner02}.  When we surmise a decoupling approximation
in evaluating the $\varphi$ and $\theta$ integrals as in
Eq.~\eqref{eq:G-U}, repeating the same calculation leads us to the
spectral function that is similar to Eq.~\eqref{eq:spectral-function}.
The only difference is that the delta functions in
Eq.~\eqref{eq:spectral-function} now acquire finite width due to the
coupling with the reservoirs.  It corresponds to the spectral function
of the Coulomb blockade regime~\cite{Ingold92,Schoeller94}.  It was
further suggested that if one implements a self-consistent decoupling
scheme, one may well understand the Kondo physics~\cite{[{See
  }][{}]Florens02,*Florens03}.
It is interesting to see how such decoupling approximation can be
improved by taking account of the compact and
non-Hermitian nature of phase fluctuations. Our work in this direction
is underway.

\section{Summary}

To summarize, we have demonstrated how one can evaluate the
coherent-state path integrals for locally interacting systems,
following its standard definition and bewaring of the operator
ordering subtlety.  The results agree with the ones by the operator
method.  In the process of calculating, we find that locally
interacting systems is equivalent to certain free-particle models
embellished with dynamical phase as well as static random variables.
Since we can view locally interacting models the strong interaction
limit of a wide-range of strongly correlated materials, it is hoped,
we use such free theories as an alternative yet viable simple
description for strongly correlated materials.

\begin{acknowledgments}
  The author gratefully acknowledges financial support from
  Grant-in-Aid for Scientific Research (C) No.~26400382
  from MEXT, Japan.
\end{acknowledgments}

\appendix




\section{Calculation via the operator method}
\label{app:operator-method}

\subsection{Grand partition function}

Since the effect of the interaction is to increase the energy by $U N(N-1)/2$ for fixed-$N$ states, we can express the grand partition function of the Hamiltonian~\eqref{eq:Hamiltonian} as
\begin{align}
& \Xi_{U} (\mu) = \Tr \left[ e^{-\beta (\hat{H} - \mu \hat{N})} \right]
= \sum_{N=0}^{\infty} Z_{N}\, e^{\beta \mu N -\beta
  \frac{U}{2} N(N-1)},
\label{eq:Xi-U}
\end{align}
where $Z_{N}$ is the canonical partition
function of the noninteracting system, defined by 
\begin{align}
& \Xi_{0}(\mu) = \sum_{N=0}^{\infty} Z_{N}\, e^{\beta \mu N}
= \prod_{\alpha} \left[ 1 \mp e^{-\beta(\epsilon_{\alpha}-\mu)}
  \right]^{\mp 1}.  
\label{eq:Xi0}
\end{align}
The sign $\mp 1$ refers to bosonic or fermionic systems.  One can write the explicit form of $Z_{N}$ via the inverse transformation of the
above as
\begin{align}
& Z_{N} = \int^{2\pi}_{0} \frac{d\theta}{2\pi}\, e^{-iN\theta}\, \Xi_{0}
  (\mu=i\theta). 
\end{align}

\subsection{Green functions}

We can solve exactly various one-particle Green functions for the
locally interacting Hamiltonian~\eqref{eq:Hamiltonian}.  The system is
not needed to be in thermal equilibrium; a generic stationary state
will suffice.
A quick way to proceed is to examine the equation of motion for
a field  operator $\psi_{\alpha}$:
\begin{align}
& i\hbar \frac{\partial \psi_{\alpha}(t)}{\partial t} 
= \left( \epsilon_{\alpha} + U \hat{N} \right) \psi_{\alpha}(t),
\end{align}
which is true for either bosonic or fermionic systems. 
We can immediately solve its time-evolution as
\begin{align}
& \psi_{\alpha}(t) = e^{-\frac{i}{\hbar} (\epsilon_{\alpha} + U
  \hat{N})t} \psi_{\alpha}
= \psi_{\alpha} \, e^{-\frac{i}{\hbar} [\epsilon_{\alpha} + U
  (\hat{N}-1)]t}.
\end{align}
With this property, we can calculate various Green functions.
For instance, the lesser and greater Green functions are found to be
\begin{align}
& G^{<}_{\alpha}(t,0) 
= \pm \frac{1}{i\hbar} \left\langle \hat{n}_{\alpha} \,
  e^{-\frac{i}{\hbar} [\epsilon_{\alpha} + U(\hat{N}-1)] t}
\right\rangle,
\label{eq:G-lesser-operator} \\
& G^{>}_{\alpha}(t,0) 
= \frac{1}{i\hbar} \left\langle e^{-\frac{i}{\hbar} (\epsilon_{\alpha}
  + U \hat{N})t} \left( 1 \pm \hat{n}_{\alpha} \right)
\label{eq:G-greater-operator}
\right\rangle,
\end{align} 
where the average $\langle \cdots \rangle$ refers to some stationary
state average.  In the energy space, they become
\begin{align}
& G_{\alpha}^{<}(\varepsilon) = \mp 2 i \pi \left\langle
  \hat{n}_{\alpha}\, \delta(\varepsilon - \epsilon_{\alpha} - U
  (\hat{N}-1) \right\rangle, \\
& G_{\alpha}^{>}(\varepsilon) = -2i\pi \left\langle (1\pm
  \hat{n}_{\alpha}) \, \delta(\varepsilon - \epsilon_{\alpha} - U
  \hat{N})  \right\rangle. 
\end{align}
We can construct all other Green functions using the results of
$G_{\alpha}^{<,>}$.  The spectral function
$\rho_{\alpha}(\varepsilon) = - \Im G_{\alpha}^{R}(\varepsilon)/\pi$
is found to be
\begin{align}
& \rho_{\alpha} (\varepsilon) 
= \Big \langle (1 \pm \hat{n}_{\alpha}) \, \delta(\varepsilon -
  \epsilon_{\alpha} - U \hat{N}) 
\notag \\ & \qquad \qquad
\mp \hat{n}_{\alpha} \,
  \delta(\varepsilon - \epsilon_{\alpha} - U(\hat{N}-1))
\Big\rangle. 
\end{align}

For fermionic systems, the results take particularly simple forms
resembling the free-particle, by the property
$\hat{n}_{\alpha}^{2} = \hat{n}_{\alpha}$. Indeed, the spectral
function becomes
\begin{align}
& \rho_{\alpha}(\varepsilon) = \left\langle \delta(\varepsilon -
  \epsilon_{\alpha} - U \hat{N}'_{\alpha}) \right\rangle,
\label{eq:spectral-function-operator}
\end{align}
with introducing $\hat{N}'_{\alpha} = \hat{N} -\hat{n}_{\alpha}$.  All
Green functions likewise have free-fermion forms only with replacing
$\epsilon_{\alpha} \mapsto \epsilon_{\alpha} + U \hat{N}'_{\alpha}$.
When we further assume that the system is in thermal equilibrium with
$\mu$ and $\beta$, the Kubo-Martin-Siggia relation makes the average
occupation number be characterized by 
the Fermi-Dirac distribution as
\begin{align}
  & \left\langle \hat{n}_{\alpha} \right\rangle 
    = \left\langle \frac{1}{e^{\beta (\epsilon_{\alpha} + U \hat{N}'_{\alpha} - \mu)} + 1 } \right\rangle,
\end{align} 
though local interaction makes it considerably deviate
from the Fermi-Dirac function regarding $\epsilon_{\alpha}-\mu$.

\section{Derivations of Eqs.~\eqref{eq:Xi-identity}--\eqref{eq:thermal-HS-decoupling}}
\label{app:detailed-derivation}

In this appendix, we present the step-by-step derivations of
Eqs.~\eqref{eq:Xi-identity}--\eqref{eq:thermal-HS-decoupling} in the
main text.  We start with the Gaussian integral formula
\begin{align}
& e^{-\beta \frac{U}{2} N^{2}}
= \int d[\tilde{\varphi}]\,
  e^{-\beta \frac{\tilde{\varphi}^{2}}{2U} - i\beta \tilde{\varphi}
  N}, 
\label{eq:simple-Gaussian}
\end{align}
where $d[\tilde{\varphi}]$ include the normalization factor and $N$ is
just a number. By using the above and Eqs.~\eqref{eq:Xi-U}--\eqref{eq:Xi0},
we immediately prove Eq.~\eqref{eq:Xi-identity} as
\begin{subequations}
\begin{align}
& \int^{\infty}_{-\infty}\!\! d[\tilde{\varphi}]\, e^{-\beta
  \frac{\tilde{\varphi}^{2}}{2U}}\, 
\Xi_{0}(\mu + \tfrac{U}{2} - i\tilde{\varphi})
\notag \\ & \quad
= \int^{\infty}_{-\infty}\!\! d[\tilde{\varphi}]\, e^{-\beta
  \frac{\tilde{\varphi}^{2}}{2U}}\, 
\sum_{N=0}^{\infty} Z_{N} e^{\beta (\mu +U/2-i\tilde{\varphi}) N}
, \\ & \quad
= \sum_{N=0}^{\infty} Z_{N} e^{\beta(\mu+U/2) N -\beta U N^{2}/2}
= \Xi_{U}(\mu).
\end{align}
\label{eq:XiU-Xi0}
\end{subequations}

We can extend the Gaussian formula~\eqref{eq:simple-Gaussian} to
the operator identity by inserting the complete basis of the occupation
number representation $|\{n_{\alpha}\} \rangle$ with the total number 
$N=\sum_{\alpha} n_{\alpha}$: 
\begin{subequations}
\begin{align}
&  e^{-\beta \frac{U}{2} \hat{N}^{2}}
= \sum_{\{ n_{\alpha}\}} |\{ n_{\alpha} \} \rangle\, 
e^{-\beta \frac{U}{2} N^{2}} \langle \{
  n_{\alpha} \}|
, \\ & \quad
= \sum_{\{ n_{\alpha}\}} |\{ n_{\alpha} \} \rangle 
\int d[\tilde{\varphi}]\,
  e^{-\beta \frac{\tilde{\varphi}^{2}}{2U} - i\beta \tilde{\varphi}
  N} \langle \{
  n_{\alpha} \}|
, \\ & \quad
= \int d[\tilde{\varphi}]\,
  e^{-\beta \frac{\tilde{\varphi}^{2}}{2U} - i\beta \tilde{\varphi}
  \hat{N}}.
\end{align}
\label{eq:op-identity-detail}
\end{subequations}
This proves the operator identity~\eqref{eq:op-identity} in the
text. 
With this identity, we can rewrite the operator $e^{-\beta \hat{H}}$ as
\begin{align}
& e^{-\beta \hat{H}} = 
\int d[\tilde{\varphi}]\,
  e^{-\beta \frac{\tilde{\varphi}^{2}}{2U} 
 -\beta \sum_{\alpha} (\epsilon_{\alpha} +i\tilde{\varphi}
                -U/2) \hat{n}_{\alpha} }.
\label{eq:exp-H}
\end{align}

We now represent both sides of Eq.~\eqref{eq:exp-H} to establish the
modification of the Hubbard-Stratonovich transformation in the
coherent-state path integral. Since the Hamiltonian $\hat{H}$ is
normal-ordered, the left-hand side of Eq.~\eqref{eq:exp-H}
is simply represented as
\begin{align}
  & \text{(LHS)} = \int \mathcal{D}[\psi,\bar{\psi}]\,
    e^{-\mathcal{S}/\hbar}, \\
  & \mathcal{S} = \sum_{\alpha,\beta}\int^{\beta\hbar}_{0} d\tau\,
\bar{\psi}_{\alpha} \left[ \left( \hbar  \partial_{\tau} 
    + \epsilon_{\alpha} \right) \delta_{\alpha \beta} +
    \frac{U}{2} \bar{\psi}_{\beta} \psi_{\beta} \right]
    \psi_{\alpha}.
\end{align}
Now we can express the right-hand side of
Eq.~\eqref{eq:exp-H} as
\begin{align}
  & \text{(RHS)} = \int d[\tilde{\varphi}] \int
    \mathcal{D}[\psi,\bar{\psi}] \, e^{-
    \frac{\beta}{2U} \tilde{\varphi}^{2} - \mathcal{S}_{e}/\hbar}
    , \\ & \quad
           = \int \mathcal{D}[\theta] \int d[\tilde{\varphi}] \int
           \mathcal{D}[\psi,\bar{\psi}] \,
           e^{- \frac{\beta}{2U}\tilde{\varphi}^{2} -\mathcal{S}_{1}/\hbar -
           \mathcal{S}_{\theta}/\hbar}. 
\label{eq:RHS2}
\end{align}
Here the Euclidean action Lagrangian $\mathcal{S}_{1}$ and
$\mathcal{S}_{\theta}$ are defined as
\begin{align}
& \mathcal{S}_{1}
= \int^{\beta\hbar}_{0} d\tau \sum_{\alpha}
                \bar{\psi}_{\alpha} \left( \hbar \partial_{\tau} +
                \epsilon_{\alpha} - \frac{U}{2} + i\tilde{\varphi} 
                \right) \psi_{\alpha}, \\
& \mathcal{S}_{\theta} = \int^{\beta\hbar}_{0}d\tau\,
                                            \frac{\hbar^{2}}{2U}
                                            (\partial_{\tau}
                                            \theta)^{2}, 
\end{align}
and, on Eq.~\eqref{eq:RHS2}, we have inserted the path integral over
bosonic field $\theta$ that satisfies the periodic boundary condition, 
\begin{align}
& \int \mathcal{D}[\theta]\, e^{- \mathcal{S}_{\theta}/\hbar}
= 1.
\end{align}
Next, we introduce a new (dynamical) field
$\tilde{\phi}(\tau) = \tilde{\varphi} - \hbar \partial_{\tau}
\theta(\tau)$ to combine $\tilde{\varphi}$ and $\theta$, and redefine
field $\psi_{\alpha}$ to absorb the phase factor. This is the reverse
manipulation of the gauge transformation in
\cite{Kamenev96,Kleinert97,Florens02,Florens03,Sedlmayr06}, with the
corresponding Jacobian
$\mathcal{D}[\theta]d[\tilde{\varphi}] = \mathcal{D}[\tilde{\phi}]$.
It enables us to express the right-hand side of Eq.~\eqref{eq:exp-H} as
\begin{align}
& \text{(RHS)}
= \int \mathcal{D}[\tilde{\phi}] \mathcal{D}[\psi,\bar{\psi}] \, 
e^{-\mathcal{S}_{e}/\hbar - \mathcal{S}_{\phi}/\hbar},
\end{align}
where $\mathcal{S}_{e}$ and $\mathcal{S}_{\phi}$ are defined in
Eqs.~(\ref{eq:thermal-HS-decoupling}c,d); this proves
Eqs.~(\ref{eq:thermal-HS-decoupling}a--d) in the text.

\section{Subtlety of the Hubbard-Stratonovich decoupling in the
  continuous time formulation}
\label{app:subtlety-of-HS}

We explicitly point out where matters the subtlety of the
Hubbard-Stratonovich transformation in the continuous time
formulation. Below we write for the one-site bosonic system but the
same argument applies equally to multi-level extension as well as
fermionic systems.

We examine how one can evaluate the matrix element
$\langle z | e^{-\frac{i t}{\hbar} \frac{U}{2} \hat{n}^{2}} |
w\rangle$ regarding the bosonic coherent state $|z\rangle = e^{\bar{z}
  b - b^{\dagger} z} |0\rangle$, with or without the
Hubbard-Stratonovich transformation. 
Direct evaluation of the matrix element leads to
\begin{align}
& \langle z | e^{-\frac{i t}{\hbar} \frac{U}{2} \hat{n}^{2}} |
w\rangle
= e^{-\frac{1}{2}(\bar{z} z + \bar{w} w)} \sum_{n=0}^{\infty}
\frac{(\bar{z} w)^{n}}{n!} e^{-\frac{i t}{\hbar} \frac{U}{2} n^{2}}.
\label{eq:matrix-element}
\end{align}
We now decompose the interaction term using the operator identity.
\begin{align}
& e^{-\frac{i t}{\hbar} \frac{U}{2} \hat{n}^{2}} =
  \int^{\infty}_{-\infty} d[\varphi]\, 
e^{\frac{i t}{\hbar} (\frac{\varphi^{2}}{2U} - \varphi \hat{n})}
= \left\langle e^{-\frac{it}{\hbar} \varphi \hat{n}} \right\rangle_{\varphi},
\end{align}
where $d[\varphi]$ includes the normalization factor and
$\langle \cdots \rangle_{\varphi}$ indicates the Gaussian average over
$\varphi$. 
One can check the correctness of this decomposition by 
putting it on the left-hand side of Eq.~\eqref{eq:matrix-element} and
using the Wick theorem with $\langle \varphi^{2}\rangle_{\varphi} =
\hbar U/(-it)$: 
\begin{align}
& \langle z|\,  \left\langle e^{-\frac{it}{\hbar} \varphi
  \hat{n}} \right\rangle_{\varphi}\, |w\rangle 
\notag \\ & \quad
= e^{-\frac{1}{2} (\bar{z}z + \bar{w} w)} \sum_{n=0}^{\infty} \frac{(\bar{z}
  w)^{n}}{n!} \exp \left[ \left\langle \tfrac{1}{2} \left(
               \tfrac{-it}{\hbar} \varphi 
               n \right)^{2}  \right\rangle_{\varphi} \right].
\label{eq:HS-decoupling-boson}
\end{align}
So far so good. Now the subtlety appears when we try to formulate it
using the path integral.  When we expand the
expression for infinitesimal time $\delta t$ up to the linear order,
we see it behave as
\begin{align}
& \langle z | e^{-\frac{i \delta t}{\hbar} \frac{U}{2} \hat{n}^{2}} |
w\rangle 
\approx e^{-\frac{1}{2}(\bar{z} z + \bar{w} w)} \sum_{n=0}^{\infty}
\frac{(\bar{z} w)^{n}}{n!} \left[ 1 - \frac{i \delta t}{\hbar}
  \frac{U}{2} n^{2}  \right].
\end{align}
Yet, this correct behavior cannot be reproduced when we truncate
Eq.~\eqref{eq:HS-decoupling-boson} up to the linear order of
$\delta t$.  The corresponding contribution comes from the
quadratic order term proportional to
$(\delta t)^{2}\langle \varphi^{2} \rangle$.
In other words, if we naively formulated the continuous-time path
integral just by expanding it regarding the linear $\delta t$ and
exponentiating it, we would get a wrong result. The missing $U/2$ term
exactly results from this slack manipulation; the use of the modified
Hubbard-Stratonovich transformation resolves the issue by avoiding such manipulation carefully.

\section{Extensions of the operative Hubbard-Stratonovich decoupling}
\label{app:extension-of-HS}

Our discussion relies on the operator version of the
Hubbard-Stratonovich transformation and its path integral
representation.  We can generalize the argument to more general forms
interaction composed by a set of mutually commuting operators, such as
$\{ \hat{n}_{\alpha}\}$.  It is because we can find the simultaneously
diagonalized basis $|\{n_{\alpha}\}\rangle$ and the operator identity
can be formulated straightforwardly [see
Eq.~\eqref{eq:op-identity-detail}].  Therefore, the
following operator identity is established:
\begin{align}
& e^{-\beta \frac{1}{2}\sum_{\alpha,\beta} U_{\alpha\beta} \hat{n}_{\alpha}
  \hat{n}_{\beta}} 
\nonumber \\ & \quad
= \int d[\tilde{\vec{\varphi}}]\,
  e^{-\frac{\beta}{2} \sum_{\alpha\beta}\tilde{\varphi}_{\alpha}
               (U^{-1})_{\alpha\beta} 
  \tilde{\varphi}_{\beta} - i\beta 
  \sum_{\alpha} \tilde{\varphi}_{\alpha} \hat{n}_{\alpha}}.
\end{align}
The path-integral representation of the Hubbard-Stratonovich
transformation should be modified accordingly to be consistent with
this operator identity.

The situation gets tricky when one treats a term involving
mutually non-commuting operators. A common example is the spin
exchange term $\hat{\vec{S}}^{2}$, which one sometimes tries to
decompose in a spin-rotational way using the Hubbard-Stratonovich
transformation.   
The decomposition relies on the 
integral identity
\begin{align}
e^{\beta J \vec{S}^{2}}
&= \int d[\vec{m}] \, e^{-\beta \frac{\vec{m}^{2}}{4J} -\beta\, \vec{m} \cdot
  \vec{S}},
\label{eq:spin-rotational-HS}
\end{align}
where $\vec{m}$ refers to a three-component vector that obeys the
Gaussian distribution respectively and $d[\vec{m}]$ includes the
normalization.  We emphasize that though the above identity is correct
for any vector $\vec{S}$, one cannot promoted it to an operator identity
with the spin operator $\hat{\vec{S}}$, because of its non-commutative
nature.  One can easily check this fact by taking the trace of both
sides of Eq.~\eqref{eq:spin-rotational-HS} for spin one-half operator ---
the left-hand side yields $2\, e^{\frac{3}{4}\beta J}$, whereas the
right-hand side, $2\, e^{\frac{\beta J}{4}} ( 1 + \beta J/2)$.
Therefore applying such types of the HS decoupling involving
non-commutative operators should be cautioned in the path
integral formulation.

\bibliographystyle{apsrev4-1} 

%

\end{document}